\documentclass[11pt,a4paper,preprintnumbers,nofootinbib]{revtex4}

\pdfoutput=1

\usepackage[T1]{fontenc}
\usepackage[utf8]{inputenc}
\usepackage{amsmath}
\usepackage{amssymb}
\usepackage{hyperref}
\usepackage[final]{graphicx}

\usepackage[english]{babel}

\usepackage{graphicx}
\usepackage[dvipsnames]{xcolor}
\usepackage{subfigure}

\usepackage{amsfonts}
\usepackage{amstext}

\usepackage{microtype}

\usepackage{multirow}
\usepackage{booktabs}

\usepackage{placeins}
\usepackage{verbatim}

\def\GeV{{\rm GeV}}
\def\MeV{{\rm MeV}}
\def\QCD{{\rm QCD}}
\def\Heff{\mathcal{H}_{\rm eff}}

\def\A{\mathcal{A}}
\def\B{\mathcal{B}}
\def\C{\mathcal{C}}

\def\O{\mathcal{O}}
\def\M{\mathcal{M}}

\def\J{\mathcal{J}}

\def\G{\mathcal{G}}

\def\Im{\mathcal{I}m}

\def\ubar{\overline{u}}
\def\dbar{\overline{d}}
\def\sbar{\overline{s}}

\def\qbar{\overline{q}}

\def\Kst{{K^\ast}}

\def\Kres{{K_{\rm res}}}

\def\Ds{{D_s}}
\def\DDs{{D_{(s)}}}

\def\Aud{{\mathcal{A}_{\rm UD}}}
\def\AudD{{\mathcal{A}_{\rm UD}^{D^+}}}
\def\AudDs{{\mathcal{A}_{\rm UD}^{D_s}}}	
\def\AudDDs{{\mathcal{A}_{\rm UD}^{D_{(s)}}}}

\begin{document}

\title{Testing the standard model with  $D_{(s)}\to K_1 (\to K\pi\pi)  \gamma$ decays  }

\author{Nico Adolph}
\email{nico.adolph@tu-dortmund.de}
\author{Gudrun Hiller}
\email{ghiller@physik.uni-dortmund.de}
\author{Andrey Tayduganov}
\email{andrey.tayduganov@tu-dortmund.de}
\affiliation{Fakult\"at Physik, TU Dortmund, Otto-Hahn-Str.4, D-44221 Dortmund, Germany}
\preprint{DO-TH 18/27}
\preprint{QFET-2018--21}

\begin{abstract}
The photon polarization in $D_{(s)}\to K_1 (\to K\pi\pi)  \gamma$ decays can be extracted from
an up-down asymmetry in the $K \pi \pi$ system, along the lines of the method known to $B \to K_1 (\to K\pi\pi)  \gamma$ decays. Charm physics is advantageous as
partner decays exist:  $D^+ \to K_1^+ (\to K\pi\pi)  \gamma$, which 
is standard model-like, and  $D_s \to K_1^+ (\to K\pi\pi)  \gamma$, which   is  sensitive to  physics beyond the standard model in $|\Delta c| =|\Delta u|=1$ transitions. The standard model predicts their  photon polarizations to be equal  up to U-spin breaking corrections, while new physics in the dipole operators can split them apart at order one level.
We estimate the  proportionality factor in the asymmetry multiplying the polarization parameter from axial vectors  $K_1(1270)$ and $K_1(1400)$  to be sizable,
up to  the few ${\cal{O}}(10)\%$  range.
The  actual value of the hadronic factor matters for the experimental sensitivity, but is not needed as an input to perform the  null test.
\end{abstract}

\maketitle

\section{Introduction}

Charm decay amplitudes are notoriously challenging  due to an often overwhelming resonance contribution in addition to poor convergence of the heavy quark expansion.
Yet, rare charm decays are of particular importance as they are sensitive to flavor and CP violation in the up-sector, complementary to $K$- and $B$-physics.
While the number of radiative and semileptonic  $|\Delta c| =|\Delta u|=1$ modes within reach of  the flavor facilities BaBar, Belle, LHCb, BESIII, and Belle II is plenty,  it needs dedicated efforts to get sufficient control over hadronic uncertainties to be able  to test the standard model (SM).  A useful strategy  known as well to the presently much more advanced  $B$-physics program is to  custom-built  observables "null tests", exploiting approximate symmetries of the SM,
such as lepton universality,  CP in $b \to s$ and $c \to u$ transitions, or $SU(3)_F$. This allows to bypass a precise, first-principle computation of hadronic matrix elements which presently may not exist.

In this work we provide a detailed study of  the up-down asymmetry ${\cal{A}}_{\rm UD}$  in the angular distributions of  $D^+\to K^+_1 (\to K \pi \pi)  \gamma$  and $D_s \to K_1^+ (\to K \pi \pi)\gamma$ decays, as a means to test the SM.
Originally proposed for $B$-decays \cite{Gronau:2001ng,Gronau:2002rz}, the method is advantageous in charm as one does not have  to rely on prior  knowledge of the $K \pi \pi$ spectrum and
theory predictions of the photon polarization. Instead, one can use the fact that the spectrum  is universal  and   the photon polarizations of   $D^+$ and $D_s$ decays in the SM are identical in the  U-spin limit  \cite{deBoer:2018zhz}.

Both $D_{(s)} \to K_1^+ \gamma$ decays are color-allowed, and are induced by $W$-exchange "weak annihilation"  (WA), which is  doubly Cabibbo-suppressed  and singly Cabbibo-suppressed in   $D^+$ and $D_s$ decays, respectively. 
 Thus, the ratio of their branching fractions $\B(D^+\to K^+_1\gamma)/\B(D_s\to K^+_1\gamma)\approx|V_{cd}/V_{cs}|^2(\tau_D/\tau_{D_s})$ is about $0.1$, taking into account the different 
  CKM elements $V_{ij}$ and life times  $\tau_{D_{(s)}}$ \cite{Tanabashi:2018oca}.
 While the $D^+$ decay is SM-like, the $D_s$ decay is a flavor changing neutral current (FCNC) process and
   is sensitive to physics beyond the SM (BSM)  in photonic dipole operators, which can alter the polarization. The photon dipole contributions in the SM are negligible due to the  Glashow-Iliopoulos-Maiani (GIM) mechanism.
 The photon polarization in the SM in $c \to u \gamma$ is predominantly left-handed, however, in the $D$-meson decays sizable hadronic corrections are expected
  \cite{Khodjamirian:1995uc,Fajfer:1998dv,Lyon:2012fk,deBoer:2018zhz}. In the proposal discussed in this work
  the polarization is extracted  from the SM-like decay $D^+ \to K_1^+ \gamma$. We test the SM by comparison  to the photon polarization in $D_s \to K_1^+ \gamma$ decays.
 Methods to look for new physics (NP) with the photon polarization in $c \to u \gamma$ transitions have been studied recently in \cite{deBoer:2018zhz,Gratrex:2018gmm}.

The plan of the paper is as follows:
General features of the decays $D^+ \to K_1^+ \gamma$ and $D_s  \to K_1^+ \gamma$ are discussed in Sec.~\ref{sec:rare}, including angular distributions for an axial-vector $K_1^+$  decaying to $K \pi \pi$.
 Predictions in the framework of QCD factorization \cite{Beneke:2000ry,Bosch:2001gv} are given, which we use  to estimate the  NP  reach.
In Sec.~\ref{sec:K1} we analyze  $K_1^+\to K^+\pi^+\pi^-$ and $K_1^+\to K^0\pi^+\pi^0$ decay chains.
Phenomenological profiles of the up-down asymmetry are worked out  in Sec.~\ref{sec:AUD}.
In Sec.~\ref{sec:con} we conclude. Auxiliary information is given in three appendices.

\section{The decays $D^+ \to K_1^+ \gamma$ and $D_s  \to K_1^+ \gamma$ \label{sec:rare}}

In Sec.~\ref{sec:angdist} we give the $\DDs \to K_1 (\to K\pi\pi) \gamma$ angular distribution that allows to probe the photon polarizations and perform the null test.
In Sec.~\ref{sec:QCD} we discuss dominant SM amplitudes and estimate the  $\DDs \to K_1(1270) \gamma$ and $\DDs \to K_1(1400) \gamma$ branching ratios. The BSM reach is investigated in
 Sec.~\ref{sec:BSM}.

\subsection{  $\DDs \to K_1 (\to K\pi\pi) \gamma$ angular distribution  \label{sec:angdist}}

The $D_{(s)}\to K_1\gamma$ decay rate, where $K_1$ is an axial-vector meson,  can be written as~\cite{deBoer:2017que} 
\begin{equation}
\Gamma^\DDs = {\alpha_e G_F^2 m_\DDs^3 \over 32 \pi^4} \left( 1 - {m_{K_1}^2 \over m_\DDs^2} \right)^3  \left( |A_L^\DDs|^2 + |A_R^\DDs|^2 \right) \,,
\label{eq:Gamma1}
\end{equation}
where $L,R$ refers to the left-handed, right-handed polarization state, respectively, of the photon. Here, $G_F$ denotes Fermi's constant and $\alpha_e$ is the fine structure constant. $A_{L,R}^{D_{(s)}}$ denote the $D_{(s)} \to K_1 \gamma$ decay amplitudes.

The polarization parameter $\lambda_\gamma^\DDs$ is  defined  as 
\begin{equation}
\lambda_\gamma^\DDs = -{1-r_\DDs^2 \over 1+r_\DDs^2}\,, \quad\quad 
r_\DDs =  \Bigg | {A_R^\DDs \over A_L^\DDs}   \Bigg |\,,
\label{eq:lambda}
\end{equation}
and can be extracted from the angular distribution in $D_{(s)}\to K_1 (\to K \pi \pi) \gamma$ decays
\begin{equation}
{\mathrm{d}^4\Gamma^\DDs \over \mathrm{d}s\mathrm{d}s_{13}\mathrm{d}s_{23}\mathrm{d}\!\cos\theta} \propto \left\{ |\J|^2(1+\cos^2\theta) + \lambda_\gamma^\DDs 2\Im[\vec{n}\cdot(\vec\J\times \vec\J^*)]\cos\theta \right\} {\rm PS}^\DDs \,,
\label{eq:MF}
\end{equation}
with the phase space factor
\begin{equation}
{\rm PS}^\DDs = {1-s/m_\DDs^2 \over 256(2\pi)^5 m_\DDs s} \,.
\label{eq:PS}
\end{equation}
Here, $s$ denotes the $K \pi \pi$ invariant mass  squared, needed for finite width effects, $ \theta$ is the angle between the normal $\vec n=(\vec p_1\times\vec p_2)/| \vec p_1\times\vec p_2|$ and the direction opposite to the photon momentum in the rest frame of the $K_1$, and $s_{ij}=(p_i+p_j)^2$ with 
four-momenta $p_i$ of the final pseudo-scalars  with
assignments specified in (\ref{eq:modeII}). Note, $p_3$ refers  to  the $K$'s momentum.  
Furthermore,  $\J$ is a helicity amplitude defined by the decay amplitude 
$A(K_1 \to K \pi \pi) \propto \varepsilon^\mu \J_\mu$ with a polarization vector $\varepsilon$ of the $K_1$, see Sec.~\ref{sec:K1} for details. $\vec \J$ are the spacial components of the four vector $\J$.
$\J$ is a feature of the resonance decay and as such  it is universal for $D^+$ and $D_s$ decays.

{}From (\ref{eq:MF}) one  can define an integrated up-down asymmetry which is proportional to the polarization parameter,
\begin{align}
 \AudDDs &=\left(\int_0^1\frac{\mathrm d^2\Gamma}{\mathrm d s \mathrm d\!\cos\theta}\mathrm d\!\cos\theta-\int_{-1}^0\frac{\mathrm d^2\Gamma}{\mathrm d s \mathrm d\!\cos\theta}\mathrm d\cos\theta\right)\bigg/\int_{-1}^1\frac{\mathrm d^2\Gamma}{\mathrm d s\mathrm d\!\cos\theta}\mathrm d\!\cos\theta\nonumber\\
 &=\frac34\frac{\left<\Im[\vec n\cdot(\vec J\times\vec J^*)]\, \kappa   \right>}{\left<|\vec J|^2\right>}\lambda_\gamma^\DDs\,,
 \label{eq:AUD}
\end{align}
where  $\kappa=\mathrm{sgn}[s_{13}-s_{23}] $ for $K^+_1 \to K^0\pi^+\pi^0$ and
$\kappa=1$ for $K_1^+ \to K^+\pi^+\pi^-$ .
 The $\left<  \, .. \, \right>$-brackets denote integration over $s_{13}$ and $s_{23}$. The reason for introducing $\kappa$ is explained in Sec \ref{sec:K1}.
The up-down asymmetry is maximal for maximally polarized photons, purely left-handed, $\lambda_\gamma^\DDs=-1$, or purely right-handed ones, $\lambda_\gamma^\DDs=+1$.

It is clear from Eqs.~\eqref{eq:MF} and~\eqref{eq:AUD} that the sensitivity to the photon polarization parameter $\lambda_\gamma^\DDs$ depends on 
$\Im[\vec{n}\cdot(\vec\J\times\vec\J^*)]$.
If this factor is zero, or too small, we have no access to $\lambda_\gamma^\DDs$. As the $\J$-amplitudes are the same for $D^+$ and $D_s$, 
the factor drops out from the ratio
\begin{equation}
{\AudD \over \AudDs} = {\lambda_\gamma^{D^+} \over \lambda_\gamma^{D_s}} = {1-r^2_{D^+} \over 1+r^2_{D^+}} {1+r^2_{D_s} \over 1-r^2_{D_s}} \,.
\label{eq:ratioAUD}
\end{equation}
In the SM, this ratio equals one in the U-spin limit. Corrections are discussed in Sec.~\ref{sec:QCD}.

In general, there is more than one $K_1$ resonance  contributing to $K \pi \pi$,  such as $K_1(1270)$ and $K_1(1400)$.
Note, the phase space suppression for the $K_J(1400)$-family and higher with respect to the $K_1(1270)$  is stronger in charm than in $B$-decays.
Therefore, a single- or double- resonance ansatz with the $K_1(1270)$ or $K_1(1400)$  is in  better shape  than in the corresponding $B \to K_1 (\to K \pi \pi) \gamma$ decays.
In the presence of more than one overlapping $K_1$ resonance, beyond the zero-width approximation, the relation between the polarization and the up-down asymmetry gets more complicated than (\ref{eq:AUD}).  The reason is that, ultimately, $r_\DDs$ and the polarization are different for $K_1(1270)$ and $K_1(1400)$, that is,  they vary with  $s$, an effect that can be controlled by cuts.
 The general formula can be seen in Appendix \ref{sec:note}.
What stays intact, however, is the SM prediction, $\left({\AudD / \AudDs}\right)_{\rm SM}=1$ up to U-spin breaking.

\subsection{SM    \label{sec:QCD}}

\begin{figure}[t!]\centering
\includegraphics[width=0.8\textwidth]{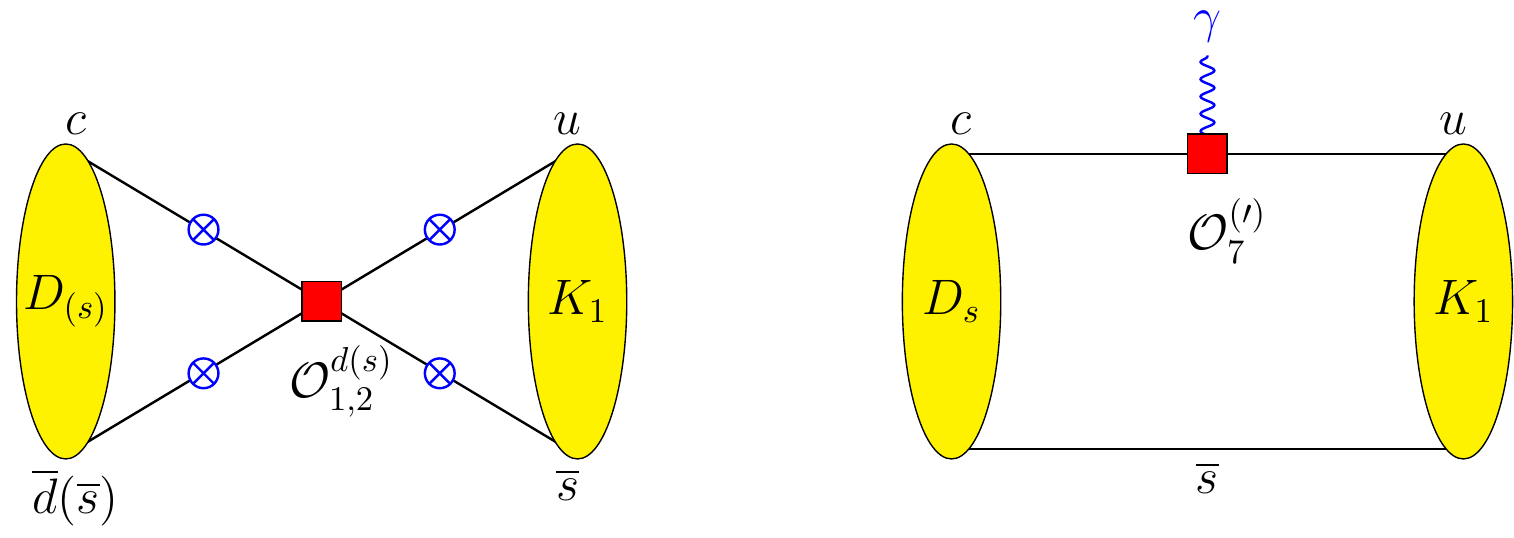}
\caption{Weak annihilation (left) and photon dipole  (right)  contributions to $D_{(s)}\to K_1\gamma$ decays. 
In the weak annihilation diagram the crosses indicate where the photon can be attached. }
\label{fig:diagrams}
\end{figure}

Rare $c\to u\gamma$ processes can be described by the effective Hamiltonian \cite{deBoer:2015boa},
\begin{equation}
\Heff = -{4G_F \over \sqrt2} \biggl[ \sum_{q=d,s} V_{cq}^*V_{us} \sum_{i=1}^2 C_i\O_i^{q} + \sum_{i=3}^6 C_i\O_i + \sum_{i=7}^8 \bigl(C_i\O_i + C_i^\prime\O_i^\prime\bigr) \biggr] \,,
\label{eq:Heff}
\end{equation}
where 
the  operators relevant to this work are defined as follows
\begin{equation}
\begin{split}
\O_1^{q=d,s} &= (\ubar_L \gamma_\mu T^a  q_L)(\qbar_L \gamma^\mu T^a  c_L) \,, \quad \O_2^{q=d,s} =(\ubar_L \gamma_\mu q_L)(\qbar_L \gamma^\mu c_L)    \, , \\
\O_7 &= {e \over 16\pi^2} m_c  \ubar_L\sigma^{\mu\nu} c_R F_{\mu\nu} \,, \quad \O_7^\prime = {e \over 16\pi^2} m_c \ubar_R \sigma^{\mu\nu} c_L F_{\mu\nu} \, ,
\end{split}
\end{equation}
with chiral left (right) projectors $L (R)$, the field strength tensor of the photon, $F_{\mu \nu}$,  and the generators of $SU(3)_c$, $T^a$, $a=1,2,3$.
Contributions to $D_{(s)}\to K_1\gamma$ decays are illustrated  in Fig.~\ref{fig:diagrams}.

In the SM both four quark operators $\O_{1,2}$ are induced at tree level, and acquire  order one coefficients at the charm quark mass $m_c$.
On the other hand, the SM contributions to the dipole operators $\O_7^{(\prime)}$ are strongly GIM-suppressed,
 $C_{7}^{\,\rm eff}\in[-1.51 - 5.51i,-0.88 - 3.25i]\times10^{-3}$ at two loop level \cite{deBoer:2017que},  and 
$C_7^\prime \sim m_u/m_c \simeq 0$.
 The $D^+ \to K_1^+ \gamma$ and $D_s  \to K_1^+ \gamma$ decays are therefore expected to be  dominated  by the four quark operators.

We employ QCD factorization methods  \cite{Bosch:2001gv} to estimate the branching ratios and the BSM sensitivity. The leading SM contribution is shown in the diagram to the left  in Fig.~\ref{fig:diagrams}, with
 the radiation of the photon from the light quark of the $D_{(s)}$ meson. The other three WA diagrams are suppressed by $\Lambda_\QCD/m_c$ and are neglected.
The corresponding WA amplitudes   for $D \to V \gamma$ have been computed in Ref.~\cite{deBoer:2017que}. We obtain~\footnote{There is a minus  sign for axial vectors relative to vector mesons from the definition of the decay constant.}  
\begin{equation}
\begin{split}
A_{L \, \rm  SM}^D  &= -  {2\pi^2 Q_d f_D f_{K_1} m_{K_1} \over  m_D  \lambda_D} V_{cd}^* V_{us} C_2   {m_D^2 \over m_D^2-m_{K_1}^2} \,, \\
A_{L \, \rm SM}^\Ds &= -  {2\pi^2 Q_d f_{D_s} f_{K_1} m_{K_1} \over m_{D_s}  \lambda_{D_s}} V_{cs}^* V_{us} C_2  {m_{D_s}^2 \over m_{D_s}^2-m_{K_1}^2} \,,
\label{eq:C7WA}
\end{split}
\end{equation}
where $Q_d=-1/3$. We also kept explicitly, {\it i.e.,} did not expand in $1/m_D$, the factors that correct for the kinematic factors in $\Gamma^\DDs$, see (\ref{eq:Gamma1}), corresponding to the matrix elements of dipole operators.
Using the  range $C_2\in[1.06,1.14]$~\cite{deBoer:2017que} we find
\begin{equation}
\begin{split}    \label{eq:BR}
\B(D^+ \to K_1^+(1270)\gamma) &= \left[ (1.3 \pm 0.3) , (1.5 \pm 0.4) \right] \times 10^{-5} \left({0.1~\GeV \over \lambda_D}\right)^2 \,, \\
\B(D^+ \to K_1^+(1400)\gamma) &= \left[ (1.4 \pm 0.6) , (1.6 \pm 0.7) \right] \times 10^{-5} \left({0.1~\GeV \over \lambda_D}\right)^2 \,, \\
\B(D_s \to K_1^+(1270)\gamma) &= \left[ (1.9 \pm 0.4) , (2.2 \pm 0.5) \right] \times 10^{-4} \left({0.1~\GeV \over \lambda_{D_s}}\right)^2 \,, \\
\B(D_s \to K_1^+(1400)\gamma) &= \left[ (2.0 \pm 0.9) , (2.4 \pm 1.0) \right] \times 10^{-4} \left({0.1~\GeV \over \lambda_{D_s}}\right)^2 \,,
\end{split}
\end{equation}
where the first (second) value corresponds to the  lower (upper)  end  of the range for the Wilson coefficient $C_2$. In each case, parametric uncertainties from the $K_1$ decay constants \eqref{eq:fK1_LCSR},
$D_{(s)}$ decay constants from  lattice-QCD $f_D=(212.15\pm1.45)~\MeV$  and $f_\Ds=(248.83\pm1.27)~\MeV$~\cite{Aoki:2016frl}, masses, life times \cite{Tanabashi:2018oca} and CKM elements \cite{Bona:2006ah}
are taken into account and added in quadrature.
The parameter  $\lambda_{D_{(s)}} \sim\Lambda_{\rm QCD}$  is poorly known, and constitutes a major uncertainty to the SM  predictions (\ref{eq:BR}).
Data on $D\to V \gamma$ branching ratios suggest a rather low value for $\lambda_D$  \cite{deBoer:2017que}.
We  use 0.1 GeV as benchmark value  for both $D$ and $D_s$ mesons.

Despite its V-A structure in the SM contributions to right-handed photons are expected, which we denote by $A_{R \, \rm SM}^\DDs$.
One possible mechanism  responsible for  $\lambda_\gamma^\DDs \neq -1$ is  a quark loop with an $\O_{1,2}$ insertion and the photon and a soft gluon attached \cite{Grinstein:2004uu}, at least perturbatively also subject to GIM-suppression \cite{deBoer:2017que}.
Here we do not need to  attempt an estimate of such effects as we take the SM fraction of right- to left-handed photons from a measurement  of $\AudD$ in  $D^+ \to K_1^+ \gamma$ decays, which has no FCNC-contribution.
(We neglect BSM effects in four quark operators.)

U-spin breaking between $D$ and $D_s$ meson decays  can  split  the photon polarizations in the SM.
While obvious sources such as phase space and CKM factors can be taken into account in a straight-forward manner,  there are further effects induced by hadronic physics.
Examples for parametric input are the decay constants, and $\lambda_\DDs$, as in (\ref{eq:C7WA}). The former has known $U$-spin splitting of $\sim 0.15$  \cite{Aoki:2016frl}, and for the latter, as not much is known, 
we assume that  the spectator quark flavor does not matter beyond that. A measurement of $D_s \to  \rho^+ \gamma$, which is a Cabibbo and  color-allowed SM-like mode
with branching ratios of order $10^{-3}$ \cite{deBoer:2017que}
can put this to a test. Nominal U-spin breaking in charm is ${\cal{O}}(0.2-0.3)$, {\it e.g.} \cite{Brod:2012ud,Hiller:2012xm,Muller:2015lua}, however, the situation for the photon polarization  is  favorable, as
only the residual breaking on the ratio of left-handed to right-handed amplitude is relevant for the null test. In the BSM study we work with U-spin breaking between  $r_{D+}$ and $r_{Ds}$
within $\pm 20 \%$ .

\subsection{BSM \label{sec:BSM}}

Beyond the SM, the GIM suppression does not have to be at work in general and the dipole coefficients can be significantly enhanced. Model-independently, the following constraints hold
\begin{align}   \label{eq:BSM}
|C_7|, |C_7^\prime| \lesssim 0.5 \, , 
\end{align}
obtained from $D \to \rho^0 \gamma$ decays~\cite{Abdesselam:2016yvr,deBoer:2017que}, and consistent with limits from  $D \to \pi^+ \mu \mu$ decays \cite{deBoer:2015boa}.

The corresponding NP contributions to $D_s  \to K_1^+ \gamma$ decays are  given as
\begin{equation}
A_{L \, \rm NP}^\Ds = m_c C_7 T^{K_1}, \quad
A_{R \,  \rm NP}^\Ds = m_c C_7^{\prime} T^{K_1} \,,
\label{eq:AL-AR}
\end{equation}
where $T^{K_1}=T_1^{D_s \to K_1}(0)$ is the form factor for the $\Ds \to K_1$ transition, defined in Appendix \ref{app:amplitudes}.  

{}From radiative $B$-decay data \cite{Amhis:2016xyh}
\begin{align}
{\cal{B}}(B \to K^{0*}(892) \gamma) & = (41.7 \pm 1.2 )   \times 10^{-6} \, , \\
{\cal{B}}(B^+ \to K_1^+(1270) \gamma) & = ( 43.8^{+7.1}_{-6.3}) \times 10^{-6} \, , \\
{\cal{B}}(B^+ \to K_1^+(1400) \gamma) & =  (9.7^{+5.4}_{-3.8} )  \times 10^{-6} \, ,
\end{align}
one infers that $T_1^{B \to K_1(1400)}/T_1^{B \to K_1(1270)} \simeq 0.5$ and $T_1^{B \to K_1(1270)}/T_1^{B \to K^*(892)} \simeq 1.1$.
Using  $ T_1^{D_s \to K^*(892)}\simeq 0.7$ from a compilation in \cite{deBoer:2017que} points to   $T^{K_1(1270)} \simeq 0.8$
and $T^{K_1(1400)} \simeq 0.4$. We use $T^{K_1(1270)}=0.8$ and $m_c=1.27$ GeV to estimate  the BSM reach.

The  SM plus NP  decay amplitudes   read
\begin{equation}
A_{L/R}^{D^+} = A_{L/R \, \rm SM}^{D^+} \,,   \quad\quad 
A_{L/R}^\Ds = A_{L/R \, \rm SM}^\Ds +A_{L/R \, \rm NP}^\Ds \,, 
\end{equation}
and  
\begin{equation}
r_{D^+} =  \Bigg | {A_{R \, {\rm SM}}^{D^+} \over A_{L \, {\rm SM}}^{D^+}}   \Bigg | \,  ,\quad\quad 
r_{D_s} = \Bigg |  {  m_c T^{K_1} C_7^{\prime\,{\rm eff}} + A_{R \, {\rm SM}}^{D_s} \over   m_c T^{K_1} C_7^{\rm eff} +  A_{L \, {\rm SM}}^{D_s}}  \Bigg | \, . 
\label{eq:rDDs}
\end{equation}

\begin{figure}[t!]\centering
\includegraphics[width=0.5\textwidth]{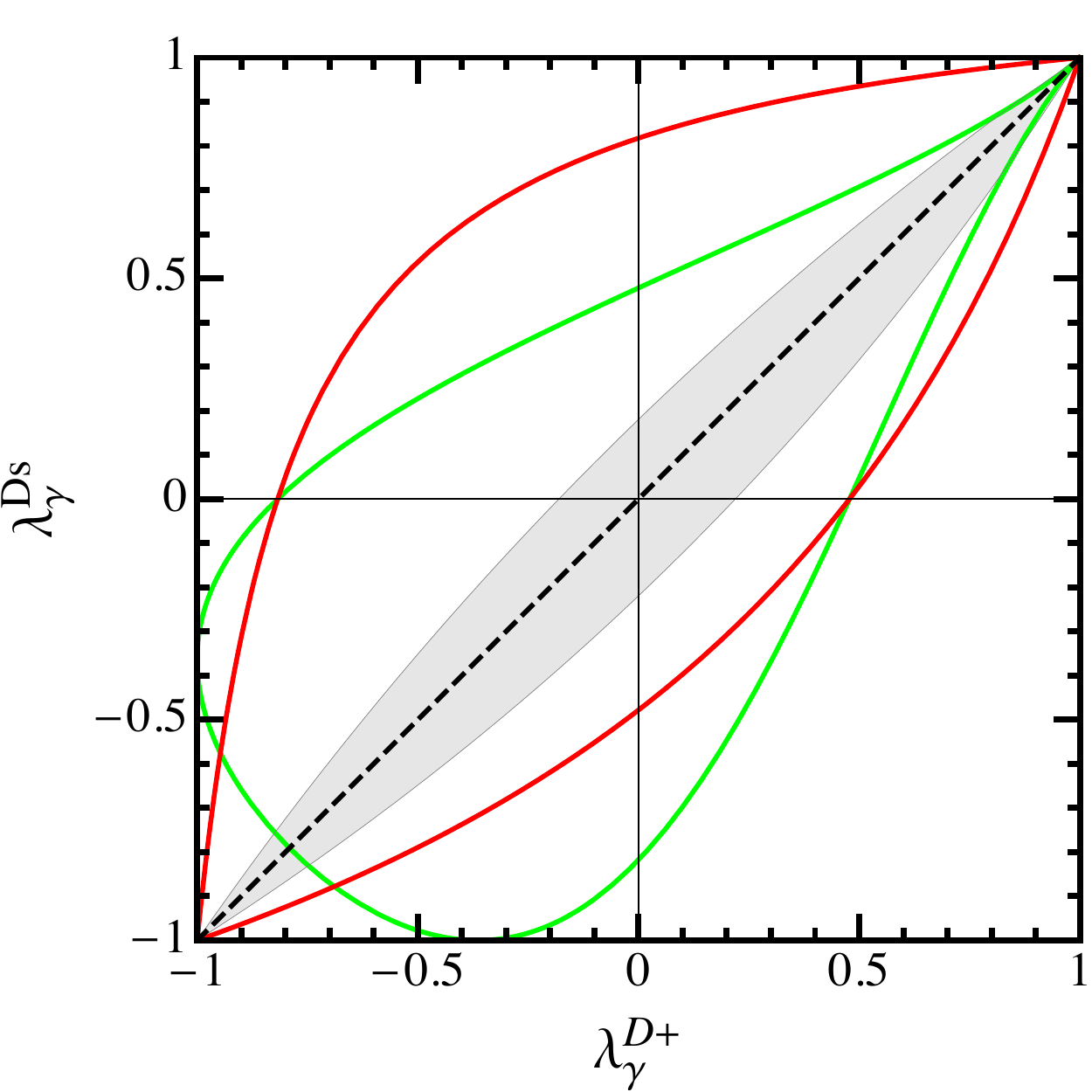}
\caption{BSM reach of $\lambda_\gamma^\Ds$ for given  $\lambda_\gamma^{D^+}$  for NP  in $C_7^\prime$ (with $C_7=0$, green curves) and NP in $C_7$ (with   $C_7^\prime=0$ red curves), within (\ref{eq:BSM})
for the $K_1(1270)$,
central values of  input,   $f_{K_1}=170$ MeV, $T^{K_1}=0.8$ and for $\lambda_{D_{(s)}}=0.1$ GeV.
 The black dashed line denotes the SM in the flavor limit, the gray shaded area illustrates  $\pm 20 \%$ U-spin breaking between $r_{D^+}$ and $r_{D_s}$.}
\label{fig:bsm}
\end{figure}

In Fig.~\ref{fig:bsm} we illustrate BSM effects that show up in $\lambda_\gamma^\Ds$ being different  from $\lambda_\gamma^{D^+}$  for NP  in $C_7^\prime$ with $C_7=0$ (green curves)  and 
in $C_7$ with $C_7^\prime=0$ (red curves), within the constraints in (\ref{eq:BSM}) for the $K_1(1270)$,
central values of  input, and for    $\lambda_{D_{(s)}}=0.1$ GeV.
We learn that NP in the left-  or right-handed dipole operator can significantly change the polarization in $D^+$ decays from the one in $D_s$ decays.
Larger  values of  $\lambda_{D_{(s)}}$ and $T^{K_1}$, and smaller values of $f_{K_1}$ enhance the BSM effects.

\section{The $K_1\to K\pi\pi$ decays  \label{sec:K1}}

Here we provide input for the $K_1\to K\pi\pi$  helicity amplitude  $\J$, which drives the sensitivity to the photon polarization in the up-down asymmetry (\ref{eq:AUD}).
After giving a general Lorentz-decomposition  we resort to a phenomenological model for the form factors, which allows us to estimate $\J$ and sensitivities.
This section is based on  corresponding studies in $B$ decays \cite{Gronau:2002rz,Tayduganov:2011ui}.
While being relevant for the sensitivity, we recall that knowledge of $\J$ in charm  is not needed as a theory input to perform the SM null test.

We consider  two $K_1$ states,  $K_1(1270) $ and $K_1(1400)$, with  spin parity $J^{\mathcal{P}}=1^+$.
For the charged resonance $K_1^+$ two types of charge combinations  exist for the final state, $K_1^+\to K^0\pi^+\pi^0$ (channel I) and $K_1^+\to K^+\pi^+\pi^-$ (channel II), 
\begin{align} 
\nonumber
{\rm I}: \quad & K_1^+(1270/1400)\to\pi\underbrace{\underbrace{^0(p_1)\pi}_{\rho^+}\overbrace{\phantom{}^+(p_2)K}^{K^{*+}}}_{K^{*0}}\phantom{}^0(p_3) \,, \\
{\rm II}: \quad & K_1^+(1270/1400)\to\pi\underbrace{\underbrace{^-(p_1)\pi}_{\rho^0}\phantom{}^+(p_2)K}_{K^{*0}}\phantom{}^+(p_3) \,, \label{eq:modeII}
\end{align}
both of which we consider in the following.

The  $K_1\to K\pi\pi$ decay amplitude can be written in terms of  the helicity amplitude $\J$ as
\begin{equation}
\M(K_{1L,R}\to K\pi\pi)^{I,II} = \varepsilon_{L,R}^\mu\J_\mu^{I,II} \,,
\end{equation}
with the $K_1$ polarization vector $\varepsilon_{L,R}^\mu=(0,\pm1,-i,0)/\sqrt2$. For a  $1^+$ state  $\J_\mu^{I,II}$ can be parameterized by two functions, $\C_{1,2}$, as 
\begin{equation} \label{eq:Jdef}
\J_\mu^{I,II} = [ \C_1^{I,II}(s,s_{13},s_{23})p_{1\mu} - \C_2^{I,II}(s,s_{13},s_{23})p_{2\mu} ] BW_{K_1}(s) \,.
\end{equation}

From here on assumptions are needed to make progress on the numerical predictions of the phenomenological profiles.
First, the $\C_{1,2}$-functions are modelled  by the quasi-two-body decays $K_1\to K\rho(\to\pi\pi)$ and $K_1\to\Kst(\to K\pi)\pi$. 
Taking into account the isospin factors for each charge mode, $K_1^+\to K^0\pi^+\pi^0$ and  $K_1^+\to K^+\pi^+\pi^-$, $\C_{1,2}^{I,II}$ can be rewritten in the following form 
\cite{Tayduganov:2011ui}
\begin{equation}
\begin{split}
\C_1^{I} &= {\sqrt2\over3}(a_{13}^\Kst-b_{13}^\Kst) + {\sqrt2\over3}b_{23}^\Kst + {1\over\sqrt3}a_{12}^{\rho} \,, \quad
\C_2^{I} = {\sqrt2\over3}b_{13}^\Kst + {\sqrt2\over3}(a_{23}^\Kst-b_{23}^\Kst) - {1\over\sqrt3}b_{12}^\rho \,, \\
\C_1^{II} &= -{2\over3}(a_{13}^\Kst-b_{13}^\Kst) - {1\over\sqrt6}a_{12}^\rho \,, \qquad\qquad\quad
\C_2^{II} = -{2\over3}b_{13}^\Kst + {1\over\sqrt6}b_{12}^\rho \,,
\end{split}
\label{eq:C12_I-II}
\end{equation}
where, using factorization,
\begin{equation}
\begin{split}
a_{ij}^V &= g_{VP_iP_j}BW_V(s_{ij})[f^V+h^V\sqrt{s}(E_i-E_j)-\Delta_{ij}] \,, \\
b_{ij}^V &= g_{VP_iP_j}BW_V(s_{ij})[-f^V+h^V\sqrt{s}(E_i-E_j)-\Delta_{ij}] \,,
\label{eq:aij-bij}
\end{split}
\end{equation}
with $\Delta_{ij}={(m_i^2-m_j^2) \over m_V^2}[f^V+h^V\sqrt{s}(E_i+E_j)]$,  $E_i={(s-s_{i3}+m_i^2) \over 2\sqrt{s}}$ and the Breit-Wigner shapes $BW_V(s_{ij})=(s_{ij}-m_V^2+im_V\Gamma_V)^{-1}$. The definitions of the form factors of the $K_1\to VP$ ($V=\Kst, \rho$ and $P=\pi , K$) decay, $ f^V, h^V$, and decay constants of the $V\to P_iP_j$ decay, $g_{VP_iP_j}$ are given in Appendix  \ref{app:formfactors}. 
The form factors are obtained in the Quark-Pair-Creation Model (QPCM) \cite{LeYaouanc:1972vsx}.

In the presence of two $K_1$ states, $K_1(1270)$ and $K_1(1400)$, this framework can be  extended  by adding  the contributions weighted by the line-shapes
\begin{equation}
\J_\mu^{I,II} = \sum_{\Kres=K_1(1270,1400)} \xi_{K_{\rm res}} \bigl[\C_{1  K_{\rm res} }^{I,II}(s,s_{13},s_{23})p_{1\mu} - \C_{2  K_{\rm res}}^{I,II}(s,s_{13},s_{23})p_{2\mu}\bigr]  BW_\Kres(s) \,,
\label{eq:J_2K1}
\end{equation}
and the parameter $\xi_{K_{\rm res}} $, which  allows to switch the states on and off individually. Importantly, in a generic  situation with all $K_1$-resonances  contributing   $\xi_{K_{\rm res}} $ takes into account the differences in their production in the weak decay.  Such effects are induced by the $K_1$-dependence of hadronic matrix elements, such as $f_{K_1} m_{K_1}$
in (\ref{eq:C7WA}), or  $T^{K_1}$ in (\ref{eq:AL-AR}). For  $f_{K_1(1400)}m_{K_1(1400)}/( f_{K_1(1270)}m_{K_1(1270)}) \sim 1.1$ and $T^{K_1(1400)}/T^{K_1(1270)} \sim 0.5$ this effect is rather mild.
The ansatz (\ref{eq:J_2K1}), which is an approximation of the  general formula (\ref{eq:AUD_general}),
 allows to compute  $\Aud/\lambda_\gamma$ as in (\ref{eq:AUD})  in  Sec.~\ref{sec:AUD} independent of the weak decays. 
Eq.~(\ref{eq:J_2K1})  becomes exact, {\it i.e.,} coincides with (\ref{eq:AUD_general}) for universal $\xi_\Kres$.

\begin{figure}[t!]\centering
\includegraphics[width=0.48\textwidth]{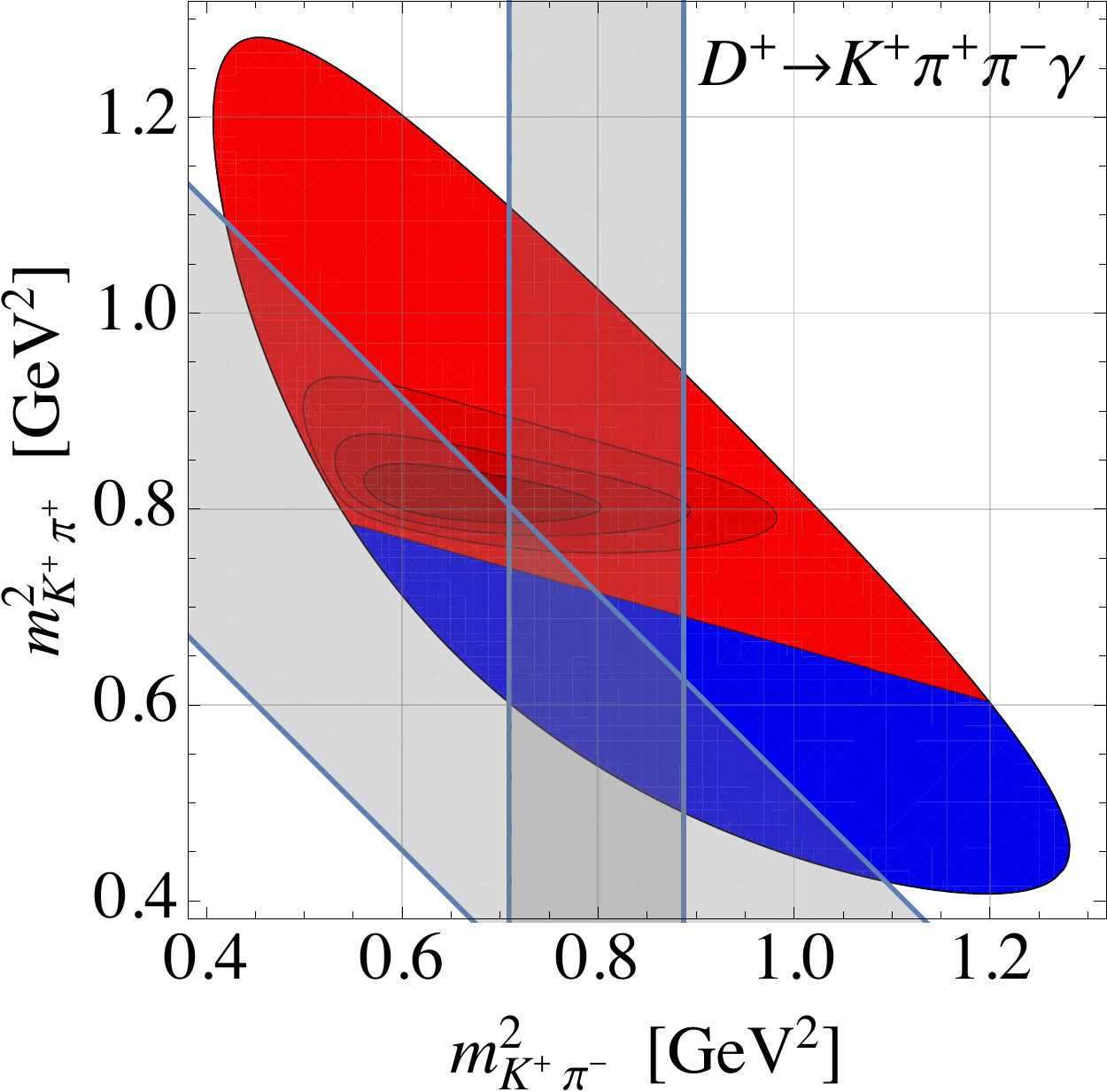}
\includegraphics[width=0.48\textwidth]{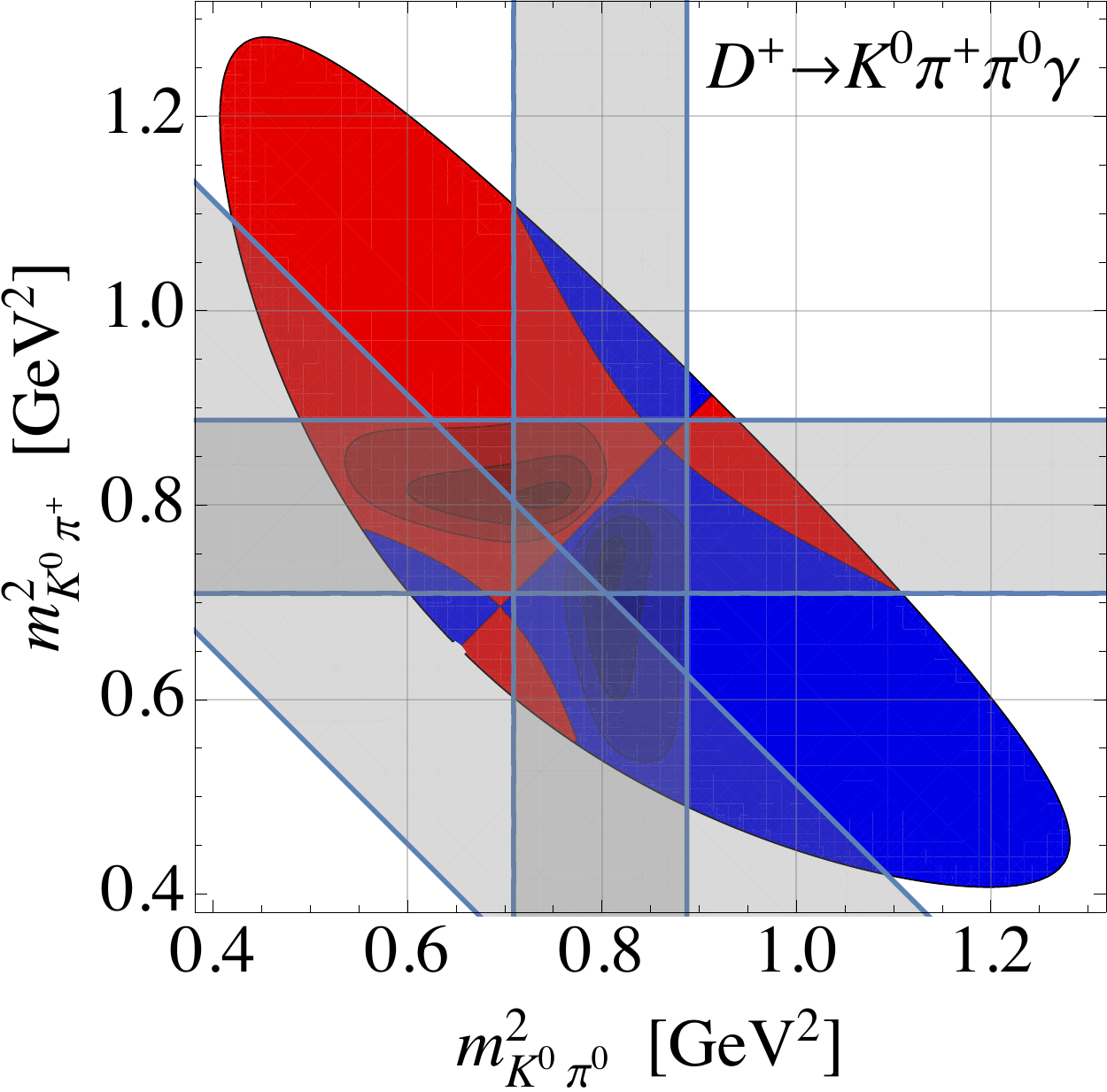}
\caption{ Dalitz contour plots of $\Im[\vec{n}\cdot (\vec\J\times\vec\J^*)]$  for $K^+\pi^+\pi^-$ (plot to the left) and $K^0\pi^+\pi^0$ (plot to the right)  at  $m_{K\pi\pi}^2=m_{K_1(1270)}^2$. Red (blue) areas correspond to positive (negative) values of $\Im[\vec{n}\cdot (\vec\J\times\vec\J^*)]$. Grey bands represent the $\Kst(\rho)$ resonance $[(m_{\Kst(\rho)}-\Gamma_{\Kst(\rho)})^2,(m_{\Kst(\rho)}+\Gamma_{\Kst(\rho)})^2]$ intervals.}
\label{fig:Dalitz_ImnJxJc}
\end{figure}

Due to isospin   $\Im[\vec{n}\cdot(\vec\J\times\vec\J^*)]$  in the $K_1^+\to K^0\pi^+\pi^0$ channel   is antisymmetric in the $(s_{13}, s_{23})$-Dalitz plane.
This can be seen explicitly by interchanging $s_{13}\leftrightarrow s_{23}$ in Eq.~\eqref{eq:C12_I-II}, which implies $\C_1\leftrightarrow\C_2$ and therefore $\Im[\vec{n}\cdot(\vec\J\times\vec\J^*)]\propto\Im[\C_1\C_2^*]$ changes  sign when crossing the $s_{13}=s_{23}$ line, see the  plot to the right in Fig.~\ref{fig:Dalitz_ImnJxJc}. Therefore, in order to have a non-zero up-down asymmetry after
$s_{13}, s_{23}$-integration, one has to define the asymmetry with  $\langle{\rm sgn}(s_{13}-s_{23})\Im[\vec{n}\cdot(\vec\J\times\vec\J^*)]\rangle$ in Eq.~\eqref{eq:AUD}. In  the $K_1^+\to K^+\pi^-\pi^+$ channel and with only one $K_1$,  the border, at which $\Aud$ changes  sign, is a straight line in the $(s_{13}, s_{23})$-plane, see the  plot to the left  in Fig.~\ref{fig:Dalitz_ImnJxJc}, which is described by  $\Im[BW_\Kst(s_{13})BW_\rho^*(s_{12})]=0$. The location of this line in the Dalitz plane depends on $s$ via
$s=s_{12}+s_{23}+s_{13} +2m_\pi^2 +m_K^2$.

\section{Up-Down Asymmetry Profiles   \label{sec:AUD}}

 In the following we work out estimates for the  up-down asymmetry  in units of the photon  polarization parameter $\A_{\rm UD}/\lambda_\gamma$, as in (\ref{eq:AUD}).
The crucial ingredient  for probing  the photon polarization is  the hadronic factor $\Im[\vec{n}\cdot (\vec\J\times\vec\J^*)] $. 
Using   (\ref{eq:J_2K1}),  and  for  two interfering resonances $a, b$, {\it  e.g.,}  $a=K_1(1270)$ and  $b=K_1(1400)$, dropping channel   $I, II$ superscripts and kinematic variables to ease notation, 
it reads
\begin{align} \nonumber
\Im[\vec{n}\cdot (\vec\J\times\vec\J^*)] = -2\Im \big [ &  \xi_a^2 \C_{1 a }\C_{2 a}^* |BW_a|^2 + \xi_b^2 \C_{1 b }\C_{2 b}^* |BW_b|^2  \\
& + \xi_a \xi_b (   C_{1 a }\C_{2 b}^*  -  C_{1 b }\C_{2 a}^*  ) BW_a BW_b^* \big  ]\,|\vec{p}_1\times \vec{p}_2|  \, , 
\label{eq:ImnJxJc}
\end{align}
which shows  the necessity of having  relative strong phases  for  a non-zero up-down asymmetry.
Such phases can come from the interference between $\Kst\pi$ and $K\rho$ channels inside of $\C_{1,2}$, as well as from the interference between the $K_1$ resonances. 
Due to the  larger number of interfering amplitudes (\ref{eq:modeII}), we  quite generally expect larger phases in  the $K_1^+\to K^0\pi^+\pi^0$ channel.
While the $K_1(1270)$ decays both to $K\rho$ and $\Kst\pi$, the $K_1(1400)$ decays predominantly to  $\Kst\pi$. We therefore expect the pure $K_1(1400)$ contribution to
 $\A_{\rm UD}/\lambda_\gamma$ in the $K^+ \pi^+ \pi^-$ channel to be very small.
 
\begin{figure}[t!]\centering
\includegraphics[width=0.49\textwidth]{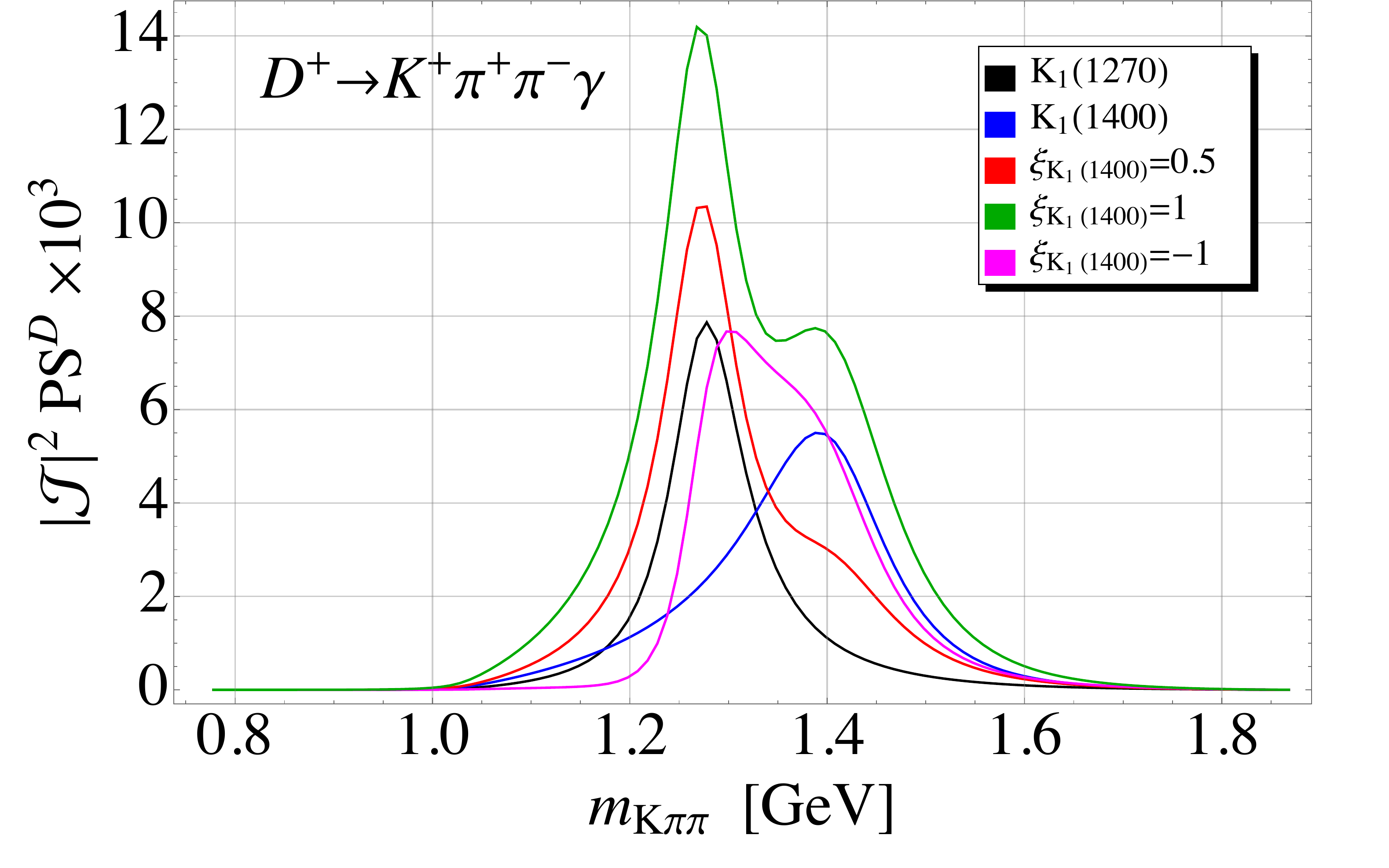}
\includegraphics[width=0.49\textwidth]{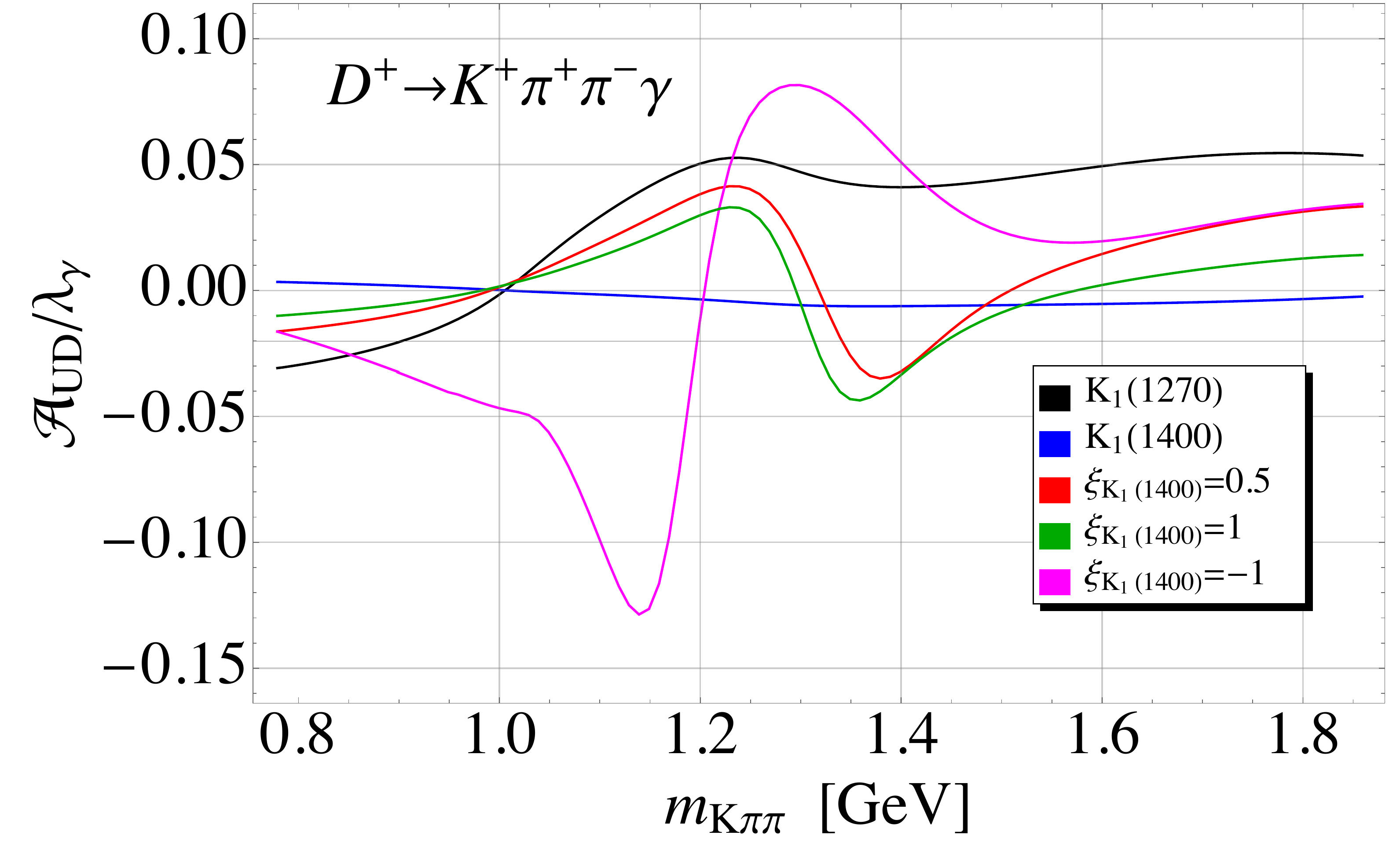}
\\
\includegraphics[width=0.49\textwidth]{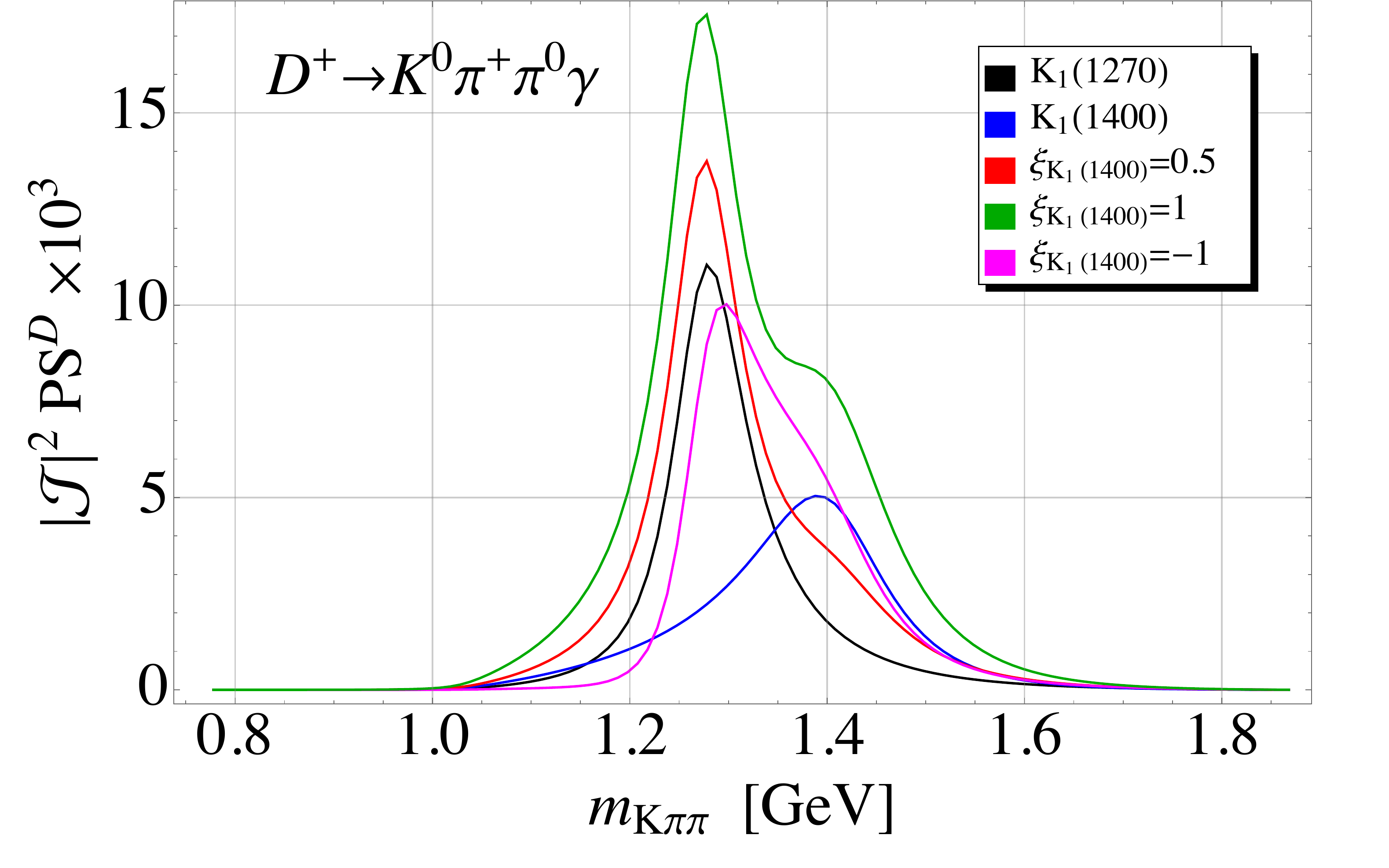}
\includegraphics[width=0.49\textwidth]{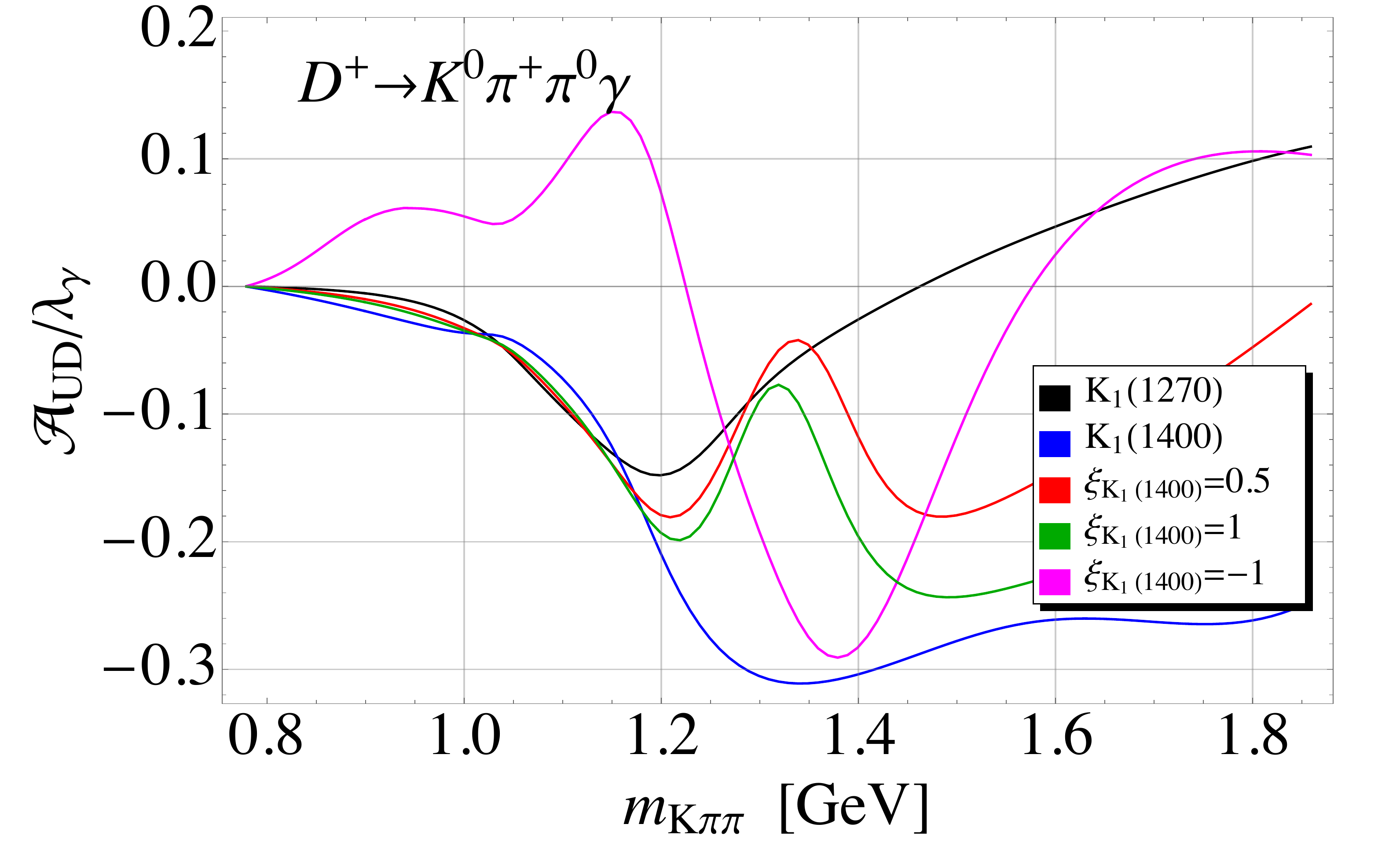}
\caption{Invariant    $K^+\pi^+\pi^-$ (upper plots) and $K^0\pi^+\pi^0$ (lower plots)  mass dependence of $|\vec\J|^2$ (plots to the left), multiplied by the four-body phase space factor \eqref{eq:PS}, and $\A_{\rm UD}/\lambda_\gamma$ (plots to the right)  for $K_1(1270,1400)$ resonances separately and with relative fraction of the $K_1(1400)$ contribution, $\xi_{K_1(1400)}$, see text for details.}
\label{fig:J2-AUD}
\end{figure}

\begin{figure}[t!]\centering
\includegraphics[width=0.49\textwidth]{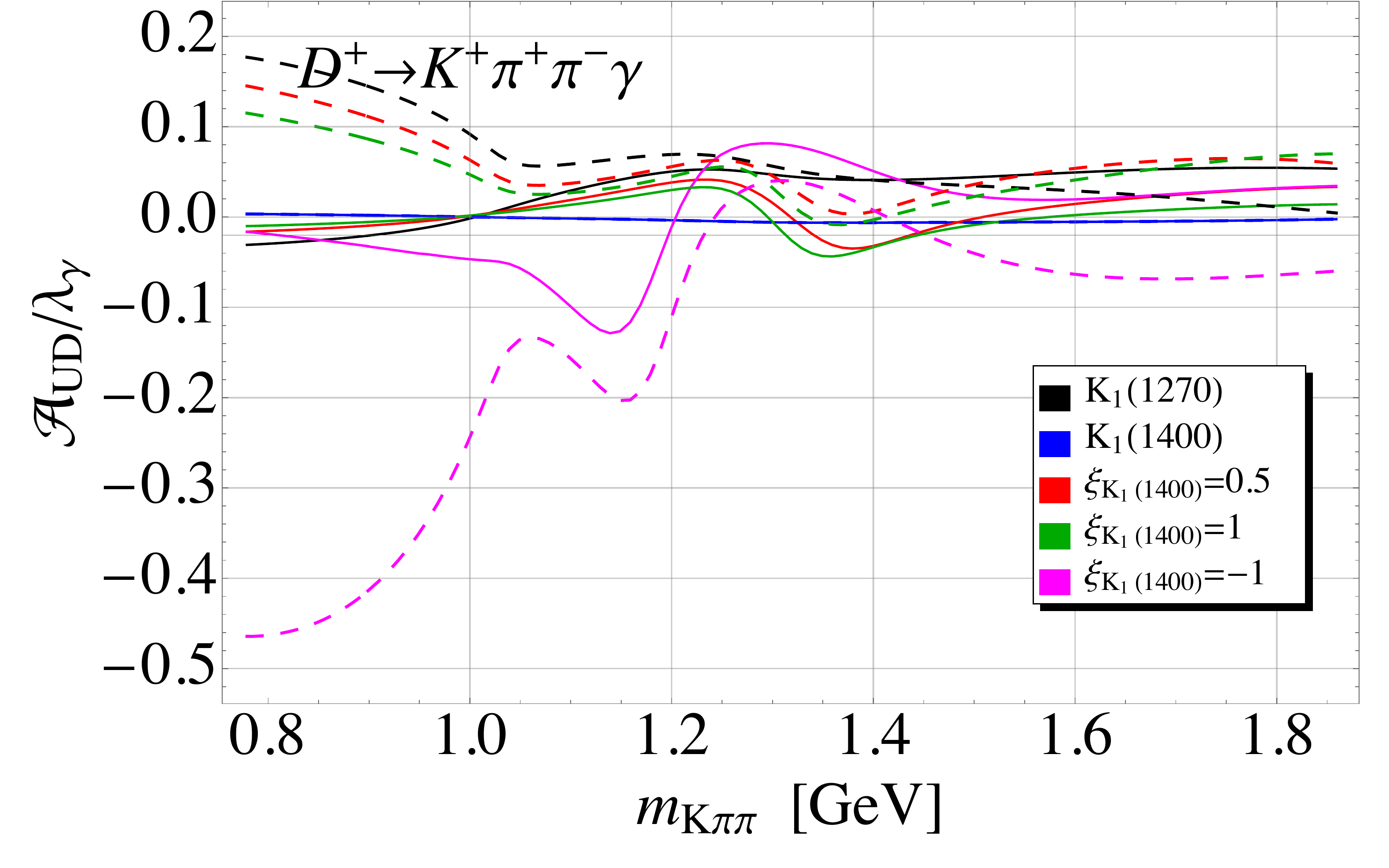}
\includegraphics[width=0.49\textwidth]{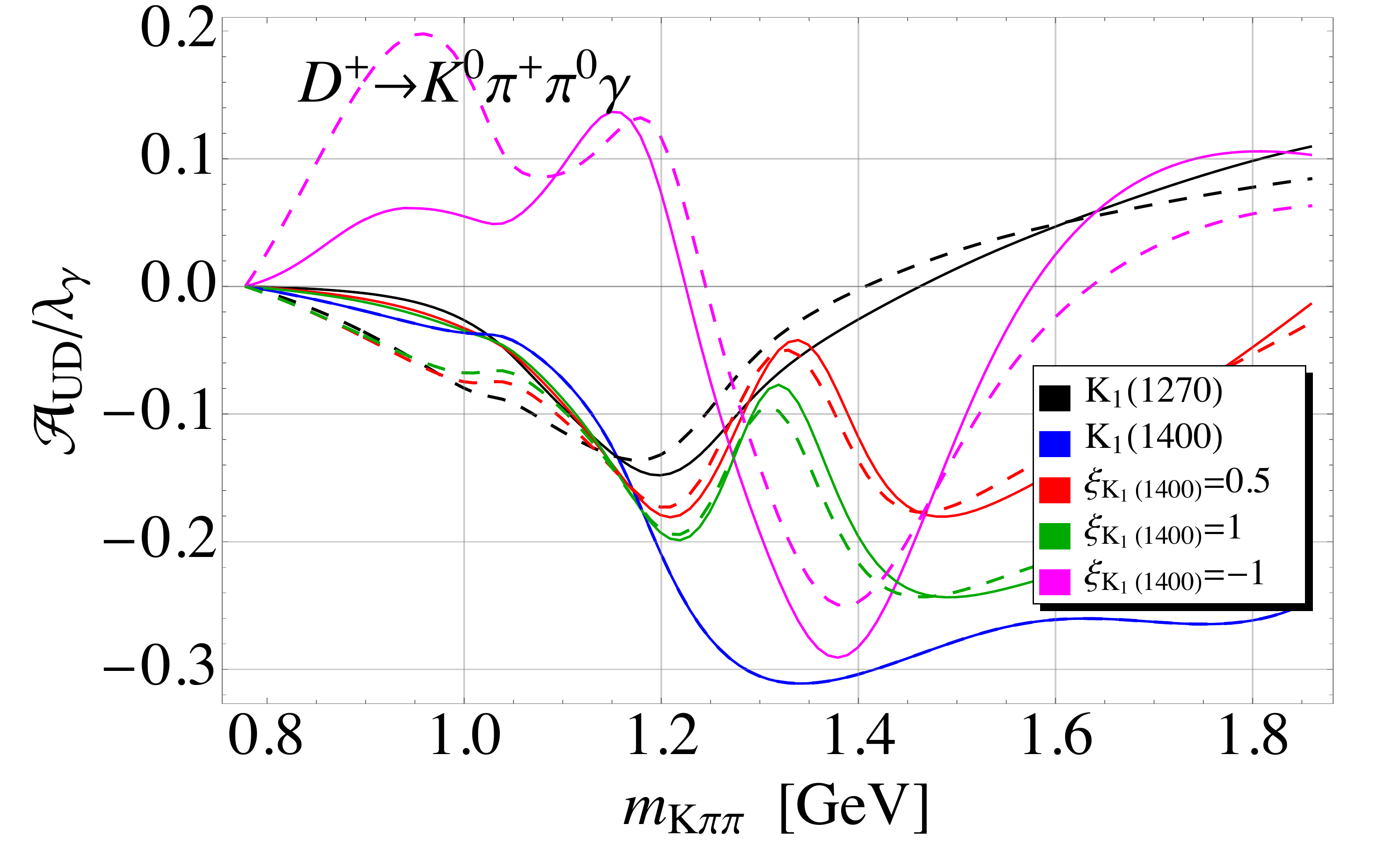}
\caption{ Invariant   $K^+\pi^+\pi^-$ (plot to the left) and $K^0\pi^+\pi^0$ (plot to the  right) mass dependence of $\A_{\rm UD}/\lambda_\gamma$ for $K_1(1270,1400)$ resonances separately and with
 relative fraction of the $K_1(1400)$ contribution, $\xi_{K_1(1400)}$. Solid lines correspond to all ``off-set'' phases equal to zero, {\it i.e.,} the  pure quark model prediction. Dashed lines represent the ``off-set'' phase $\delta_\rho={\rm arg}[\M(K_1(1270)\to K\rho)_S/\M(K_1(1270)\to(\Kst\pi)_S)]=-40^\circ$.}
\label{fig:AUD_rho40deg}
\end{figure}

\begin{figure}[t!]\centering
\includegraphics[width=0.49\textwidth]{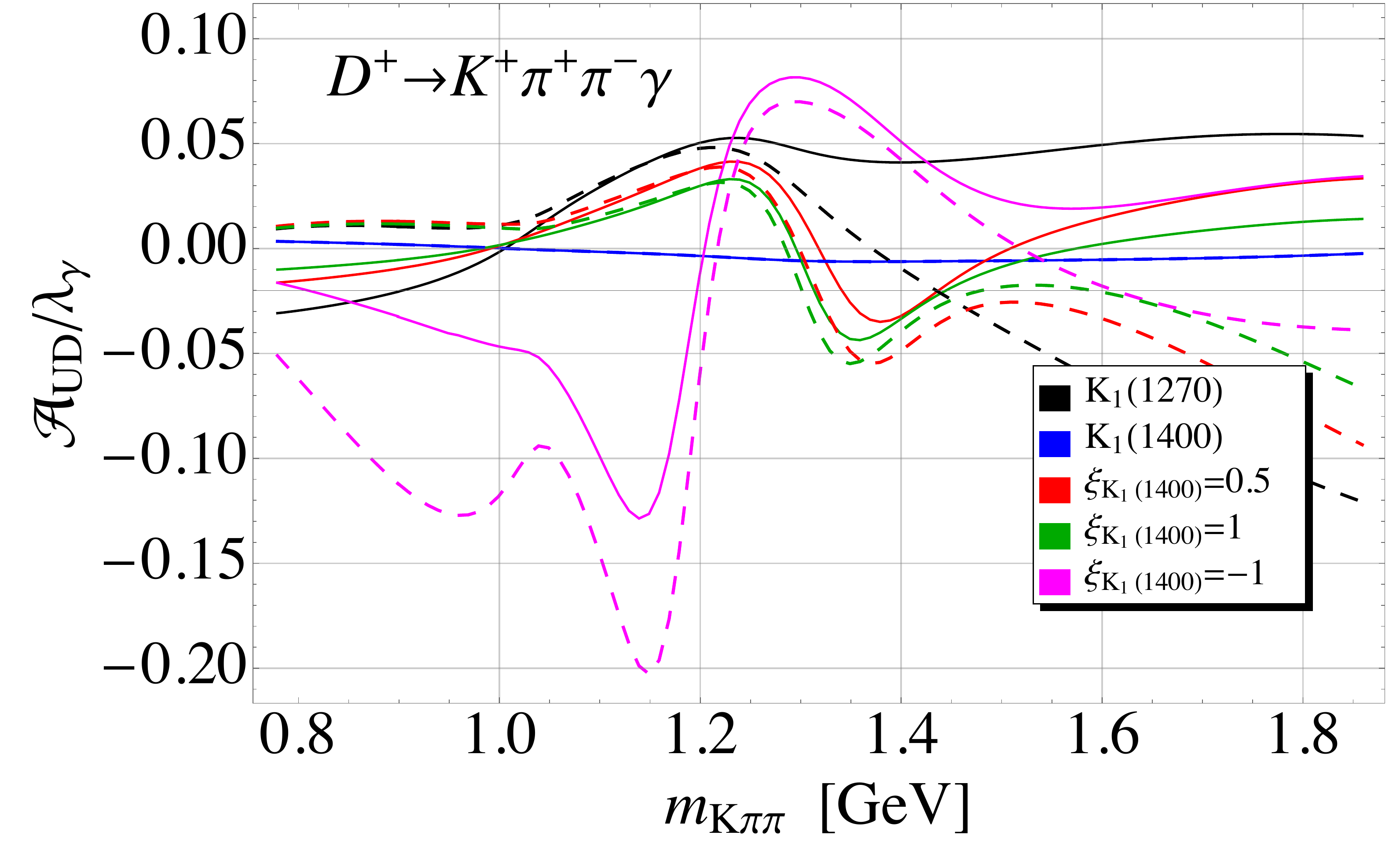}
\includegraphics[width=0.49\textwidth]{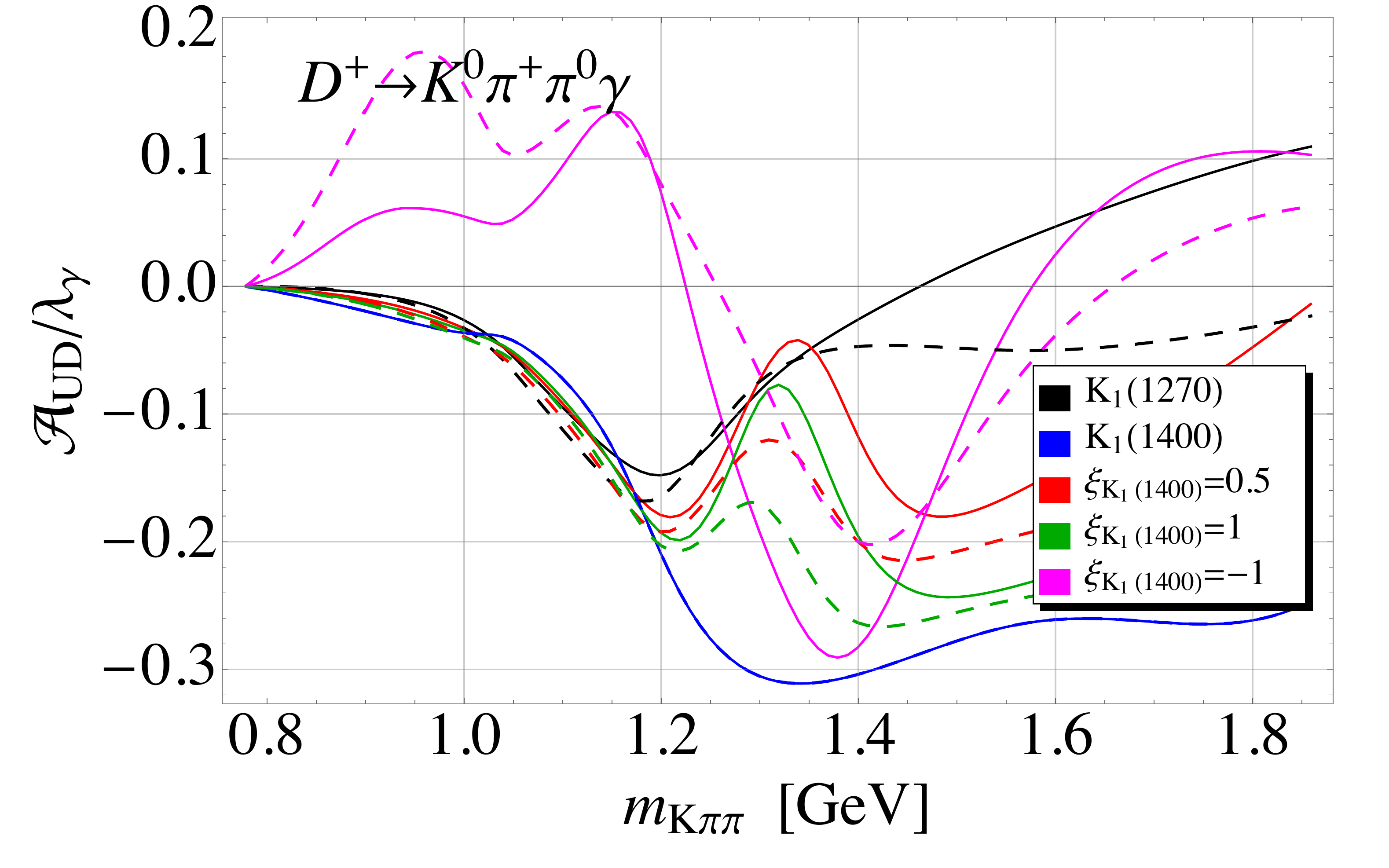}
\caption{ The same as Fig.~\ref{fig:AUD_rho40deg} for $\delta_\rho=0$ and with 
dotted lines representing the ``off-set'' phase $\delta_D={\rm arg}[\M(K_1(1270)\to (\Kst\pi)_D)/\M(K_1(1270)\to(\Kst\pi)_S)]=90^\circ$.}
\label{fig:AUD_KstD90deg}
\end{figure}

In Fig.~\ref{fig:J2-AUD} we show the $m_{K \pi \pi}$ dependence of $|\vec\J|^2$  (plots to the left) and $\A_{\rm UD}/\lambda_\gamma$  (plots to the right). The different colors refer to different ratios of the $K_1(1270)$ and $K_1(1400)$ contributions. Specifically, black,  red, green and magenta lines correspond to $\xi_{K_1(1400)}=0, +0.5, +1$ and $-1$, respectively, for fixed $\xi_{K_1(1270)}=1$. The blue curve refers to only the $K_1(1400)$ being present, with
$\xi_{K_1(1270)}=0$. Upper (lower) plots are for channel II (channel I).

The  measured  invariant mass $m_{K \pi \pi}$ spectrum in $B^+ \to K^+ \pi^+ \pi^- \gamma$ decays \cite{Aaij:2014wgo,Guler:2010if,Sanchez:2015pxu}
exhibits the dominant $K_1(1270)$-peak along with a  $K_1(1400)$-shoulder, plus higher resonances. For our model, these
 measurements suggest a value of  $\xi_{K_1(1400)}/\xi_{K_1(1270)}$ around $+1$, see Fig.~\ref{fig:J2-AUD} , consistent with expectations based on small $K_1$-dependence, see Sec. \ref{sec:K1}.
We also note that resonances higher  than the $K_1(1270)$ and the $K_1(1400)$ contribute,  such as the $K^\star_2(1430)(2^+)$ and the $\Kst(1410)(1^-)$, which are not taken into account in  our analysis.
Our predictions  therefore  oversimplify the situation  for $m_{K \pi \pi} \gtrsim 1400$ MeV.

Since  the up-down asymmetry  is  sensitive to complex phases in the $K_1$ decay amplitudes, we test several possible sources apart from the ones coming from the Breit-Wigner functions of the $K_1, K^\ast$ and the  $\rho$.
As expected, it turns out that such phases have only a negligible effect on  the $|\vec\J|^2$ distributions, and we do not show corresponding plots. 
The Belle collaboration in the analysis of $B^+\to J/\psi K^+\pi^+\pi^-$ and $B^+\to \psi^\prime K^+\pi^+\pi^-$ decays signals a non-zero phase,
\begin{equation}
\delta_\rho={\rm arg}\biggl[{ \M(K_1(1270)\to(K\rho)_S) \times \M(\rho\to\pi\pi) \over \M(K_1(1270)\to(\Kst\pi)_S) \times \M(\Kst\to K\pi) }\biggr] \,,
\end{equation}
as $\delta_\rho=-(43.8\pm4.0\pm7.3)^\circ$ \cite{Guler:2010if}. A similar value was found in the reanalysis of the ACCMOR data~\cite{Daum:1981hb} by the Babar collaboration, as $\delta_\rho=(-31\pm1)^\circ$ \cite{Aubert:2009ab}. Therefore, we add an additional phase $\delta_\rho=-40^\circ$ to the $K\rho$ $S$-wave~\footnote{Due to the smallness of the $K\rho$ $D$-wave amplitude we neglect its contribution in our study.} amplitude and consider it as theoretical uncertainty. The effect of this additional phase in $\A_{\rm UD}$ (dashed curves) in comparison with the QPCM  predictions  (solid curves) is presented in Fig.~\ref{fig:AUD_rho40deg}. We also investigate the impact  of the additional phase $\delta_D={\rm arg}[\M(K_1(1270)\to (\Kst\pi)_D)/\M(K_1(1270)\to(\Kst\pi)_S)]=90^\circ$. The result can be seen in Fig.~\ref{fig:AUD_KstD90deg}. Note that $\delta_\rho$ and $\delta_D$ vanish in the QPCM and are therefore termed ``off-set'' phases.

We learn from Figs.~\ref{fig:J2-AUD} -- \ref{fig:AUD_KstD90deg} that  $\A_{\rm UD}/\lambda_\gamma$  profiles with $\xi_{K_1(1400)} =0.5, 1$ (red,  green curves, respectively) can be of  the order $\sim 0.05-0.1$ (channel II) and $\sim 0.2-0.3$ (channel I), which are, as expected,  larger for $ K^0  \pi^+ \pi^0$ than for  $K^+ \pi^+ \pi^-$ final states. Adding phenomenological strong phases  such as $\delta_\rho$ and $\delta_D$ has a significant effect  for channel II. 
As zero-crossings can occur it may be disadvantageous to not use $m_{K \pi \pi}$ bins,  in particular, for channel II. The position of the zeros, however, cannot be  firmly predicted, although the one
 at $m_{K^+ \pi^+ \pi^-} \simeq 1 $GeV, whose origin is discussed at the end of Sec.~\ref{sec:K1},  is quite stable, as well as the one at $m_{K^+ \pi^+ \pi^-} \simeq 1.3 $GeV.
 The latter stems from  $K_1(1270)$ and $K_1(1400)$ interference.
 
 Strong phases and, related to this, $K_1$-mixing, constitute the main sources of uncertainty.
 Figs.~\ref{fig:J2-AUD} -- \ref{fig:AUD_KstD90deg} are obtained for  fixed mixing angle $\theta_{K_1}=59^\circ$, see Appendix \ref{app:formfactors}.
 Varying $\theta_{K_1}$ within its 1 $\sigma$ range, $\pm 10^\circ$,  determined within QPCM, as well as $\delta_D\in[0,2\pi]$ for $\delta_\rho=0,-40^\circ$, we find  for the $m_{K \pi \pi}$-integrated up-down asymmetry assuming $K_1(1270)$ dominance the ranges
 $[-30, +2] \, \%$ (channel I) and $[+2, +13] \, , \%$ (channel II). Recall that the latter exhibits cancellations so that locally the asymmetry can be larger.
 Our results are compatible with the findings $[-10,-7]\%$ (channel I) and $[-13,+24]\%$ (channel II) of  Ref.~\cite{Gronau:2017kyq}, which are based on $K_1(1270)$ dominance.
 Note that  Ref.~\cite{Gronau:2017kyq} uses $\kappa= {\rm sgn}(s_{13}-s_{23})$
 for both channels. Our prediction for channel II in this convention reads $[-18, +8] \, \%$.
 
We stress that the estimates are subject to sizable uncertainties and serve as a zeroth order study to explore the BSM potential in $\Ds \to K_1 \gamma$ decays.
 $K \pi \pi$ profiles from  the $B$-sector can be linked to charm physics,  and vice versa.

\section{Conclusions \label{sec:con}}

New physics may be  linked to flavor, and $K,D,$ and $B$ systems  together are required to decipher its family structure.
Irrespective of this global picture, SM tests in semileptonic and radiative $c \to u$ transitions are interesting per se, and quite unexplored territory today:
present bounds on short-distance couplings are about two orders of magnitude away from the SM \cite{deBoer:2017que,deBoer:2015boa}.

We study a null test of the SM in radiative rare charm decays based on the comparison of  the 
up-down asymmetry   in   $D^+ \to K_1^+ (\to K\pi\pi)  \gamma$, which is SM-like,  to the one in $D_s \to K_1^+ (\to K\pi\pi)  \gamma$, which is  an FCNC.
The up-down asymmetry depends on  the photon polarization, subject to BSM effects  in the   $|\Delta c| =|\Delta u|=1$ transition.

We find that, model-independently, NP  in photonic dipole  operators can alter the polarization of $D_s \to K_1^+ (\to K\pi\pi)  \gamma$ from the  SM value at order one level,
 see  Fig.~\ref{fig:bsm}.
We estimate the proportionality factor between the integrated up-down asymmetry (\ref{eq:AUD}) and the polarization parameter to be 
up to 
${\cal{O}}(5-10)\%$, and $40\%$ in extreme cases, for $K_1^+\to K^+\pi^+\pi^-$  and ${\cal{O}}(20-30)\%$ for $K_1^+\to K^0\pi^+\pi^0$, respectively,  see  Figs.~\ref{fig:J2-AUD} -- \ref{fig:AUD_KstD90deg}.
As in previous studies carried out for  $B \to  K_1^+ (\to K\pi\pi)  \gamma$ decays there  are sizable uncertainties associated with these estimates. Unlike in $B$-physics, these do not affect the SM null test.
With branching ratios (\ref{eq:BR}) of  ${\cal{B}}(D^+ \to K_1^+ \gamma)$ of  ${\cal{O}}(10^{-5})$ and ${\cal{B}}(D_s \to K_1^+ \gamma)$ of ${\cal{O}} (10^{-4})$  analyses of up-down asymmetries in charm 
constitute  an interesting NP  search for current and future flavor facilities.

\section*{Acknowledgements}

We are grateful to Stefan de Boer and Emi Kou for useful discussions and comments on the manuscript.
This work has been supported  by the DFG Research Unit FOR 1873 ``Quark Flavour Physics and Effective Field Theories''.

\appendix

\section{Matrix elements \label{app:amplitudes}}

The matrix element of the electromagnetic dipole operator  can be parametrized as
\begin{equation}
\begin{split}
\langle K_1(\varepsilon,k)|\ubar\sigma_{\mu\nu}(1\pm\gamma_5)q{^\nu} c|D_s(p)\rangle &= T_2^{K_1}(q^2) \left[ \varepsilon_\mu^* (m_\Ds^2-m_{K_1}^2) - (\varepsilon^* p)(p+k)_\mu \right] \\
&+ T_3^{K_1}(q^2) (\varepsilon^* p) \left[ q_\mu - {q^2 \over m_\Ds^2-m_{K_1}^2} (p+k)_\mu \right] \\
&\pm 2T_1^{K_1}(q^2) i\epsilon_{\mu\nu\rho\sigma} \varepsilon^{\nu*} p^\rho k^\sigma \,,
\end{split}
\label{eq:tensor_FF}
\end{equation}
with  $T_1^{K_1}(0)=T_2^{K_1}(0)$. 

The $K_1$ and $\DDs$  decay constants are defined as
\begin{align}
\langle K_1(\varepsilon,k)|\ubar\gamma_\mu\gamma_5 s|0\rangle = f_{K_1} m_{K_1} \varepsilon_\mu^*  \, , 
\label{eq:fK1} \\
\langle 0|\dbar(\sbar)\gamma_\mu\gamma_5 c|\DDs(p) \rangle = i f_\DDs p_\mu  \, . 
\label{eq:fDDs}
\end{align}

We employ the following values for the  $K_1$ decay constants
\begin{equation}
\begin{split}
f_{K_1(1270)} &= (170 \pm 20)~\MeV \,, \\
f_{K_1(1400)} &= (175 \pm 37)~\MeV \,.
\end{split}
\label{eq:fK1_LCSR}
\end{equation}
Here, $f_{K_1(1270)}$ is extracted from $\B(\tau^-\to K_1(1270)^-\nu_\tau)^{\rm exp}=(4.7\pm1.1)\times10^{-3}$ \cite{Tanabashi:2018oca}, as
\begin{equation}
\B(\tau\to K_1\nu_\tau) = \tau_\tau{G_F^2 \over 16\pi} |V_{us}|^2 f_{K_1}^2 m_\tau^3 \biggl( 1 + {2m_{K_1}^2 \over m_\tau^2} \biggr) \biggl( 1 - {m_{K_1}^2 \over m_\tau^2} \biggr)^2 \,.
\end{equation}
The value  of $f_{K_1(1270)}$ from a light cone sum rule calculation  \cite{Hatanaka:2008xj} is consistent with the data-based value  (\ref{eq:fK1_LCSR}) assuming the SM.
The value of  $f_{K_1(1400)}$  is taken from Ref.~\cite{Hatanaka:2008xj}; we added statistical and systematic uncertainties in quadrature and symmetrized the uncertainties.
$\B(\tau^-\to K_1(1400)^-\nu_\tau)^{\rm exp}=(1.7\pm2.6)\times10^{-3}$ \cite{Tanabashi:2018oca} has too large uncertainty  to allow for an extraction of $f_{K_1(1400)}$, however,
yields a 90 \% CL upper limit   as $|f_{K_1(1400)}|< 235 $ MeV, consistent with (\ref{eq:fK1_LCSR}).

\section{$K_1\to VP$ form factors   \label{app:formfactors}}

The hadronic form factors, $f_V$ and $h_V$, defined as  
\begin{equation}
\M(K_1\to VP)=\varepsilon_{K_1}^\mu(f^V g_{\mu\nu}+h^Vp_{V\mu}p_{K_1\nu})\varepsilon_V^{\nu*}
\label{eq:Amp_tensor_K1-VP}
\end{equation}
are related to the partial $S,D$ wave amplitudes,
\begin{equation}
\begin{split}
f^V &= -A_S^V-{1\over\sqrt2}A_D^V \,, \\
h^V &= {E_V\over\sqrt{s}|\vec{p}_V|^2}\left[\left(1-{\sqrt{s_V}\over E_V}\right)A_S^V+\left(1+2{\sqrt{s_V}\over E_V}\right)\frac{1}{\sqrt2}A_D^V\right] \, . 
\end{split}
\label{eq:K1-VP_form_factors}
\end{equation}
These partial wave amplitudes are computed in the framework of the $^3P_0$ QPCM \cite{LeYaouanc:1972vsx}. The details of the computation and expressions for $A_{S,D}^{\Kst/\rho}$ can be found in Ref.~\cite{Tayduganov:2011ui}.
Due to  $SU(3)$ breaking, the $K_1(1270)$ and $K_1(1400)$ mesons are an admixture of the spin singlet and triplet $P$-wave states $K_{1B}(1^1P_1)$ and $K_{1A}(1^3P_1)$, respectively,
\begin{align}
|K_1(1270)\rangle &= |K_{1A}\rangle\sin\theta_{K_1} + |K_{1B}\rangle\cos\theta_{K_1} \,, \\
|K_1(1400)\rangle &= |K_{1A}\rangle\cos\theta_{K_1} - |K_{1B}\rangle\sin\theta_{K_1} \,,
\label{eq:K1_mixing}
\end{align}
with  mixing angle  $\theta_{K_1}=(59\pm10)^\circ$ \cite{Tayduganov:2011ui}, which has been obtained  from $K_1 \to V P$ decay data.

\section{General formula for the up-down asymmetry \label{sec:note}}

The reduced amplitude of $\DDs\to \Kres \gamma \to K\pi\pi \gamma$ decays can be written as the product of the weak decay amplitude $\M_{L/R}^{\DDs,\Kres} $ and strong decay amplitude $\J_\mu^\Kres$ as 
\begin{equation}
\G_{\mu,\,L/R}^\DDs = \sum_{\Kres} \M_{L/R}^{\DDs,\Kres} \, \J_\mu^\Kres \,.
\label{eq:totalampGf}
\end{equation}
Multiplying $\G_{\mu,\,L/R}^\DDs $ by the photon polarization vector and integrating over azimuthal angles, we obtain the general formula for modulus squared of the matrix element
\begin{equation}
\overline{|\M^\DDs|^2} \propto \big(|\vec\G_L^\DDs|^2 + |\vec\G_R^\DDs|^2 \big) (1+\cos^2\theta) - 2\Im\big[ \vec{n}\cdot\big( \vec\G_L^\DDs \times \vec\G_L^{\DDs*} - \vec\G_R^\DDs \times \vec\G_R^{\DDs*} \big) \big] \cos\theta \, . 
\end{equation}
This expression holds even beyond  (\ref{eq:totalampGf}), such as for non-resonant contributions, as long as the $K \pi \pi$ system is in the same spin, parity state as $\Kres$, $1^+$.
The up-down asymmetry then reads
\begin{equation}
\AudDDs = {\displaystyle{ \biggl[ \int_0^1- \int_{-1}^0 \biggr] {\mathrm d^2\Gamma^\DDs \over \mathrm d s \mathrm d\!\cos\!\theta}\mathrm d\!\cos\!\theta} \over \displaystyle{\int_{-1}^1{\mathrm d^2\Gamma^\DDs \over \mathrm d s \mathrm d\!\cos\!\theta}\mathrm d\!\cos\!\theta}}
= -{3\over4} {\big\langle \Im\big[ \vec{n}\cdot\big( \vec\G_L^\DDs \times \vec\G_L^{\DDs*} - \vec\G_R^\DDs \times \vec\G_R^{\DDs*} \big) \big] \big\rangle \over \big\langle |\vec\G_L^\DDs|^2 + |\vec\G_R^\DDs|^2 \big\rangle} \, . 
\label{eq:AUD_general}
\end{equation}

\end{document}